\documentclass[11pt]{book}

\usepackage{amssymb}
\usepackage{amsmath}
\usepackage{amsfonts}
\usepackage{epsfig}

\makeatletter

\renewenvironment{thebibliography}[1]
     {\section*{\bibname}%
      \@mkboth{\MakeUppercase\bibname}{\MakeUppercase\bibname}%
      \list{\@biblabel{\@arabic\c@enumiv}}%
           {\settowidth\labelwidth{\@biblabel{#1}}%
            \leftmargin\labelwidth
            \advance\leftmargin\labelsep
            \@openbib@code
            \usecounter{enumiv}%
            \let\p@enumiv\@empty
            \renewcommand\theenumiv{\@arabic\c@enumiv}}%
      \sloppy
      \clubpenalty4000
      \@clubpenalty \clubpenalty
      \widowpenalty4000%
      \sfcode`\.\@m}
     {\def\@noitemerr
       {\@latex@warning{Empty `thebibliography' environment}}%
      \endlist}

\makeatother

\newcommand{\sect}[1]{\setcounter{equation}{0}\section{#1}}

\renewcommand\bibname{References}

\newcommand{\re}{\mathop{\mathrm{Re}}}
\newcommand{\im}{\mathop{\mathrm{Im}}}
\newcommand{\Heff}{\mathcal{H}_{\mathrm{eff}}}
\newcommand{\aver}[1]{\left\langle{#1}\right\rangle}
\newcommand{\T}{\mathcal{T}}

\begin{document}

\setlength{\baselineskip}{5.0mm}


\chapter[Resonance Scattering in Chaotic Systems]{Resonance Scattering of Waves in Chaotic Systems}
\thispagestyle{empty}
\ \\ %
\noindent %
{{\sc Y. V. Fyodorov}$^1$ and {\sc D. V. Savin}$^2$
\\~\\$^1$School of Mathematical Sciences,
University of Nottingham, \newline Nottingham, NG7 2RD, UK
\\
$^2$Department of Mathematical Sciences,
Brunel University West London,\newline Uxbridge UB8 3PH, UK}

\setlength{\unitlength}{1cm}
\begin{picture}(10,0)(-3,-11.5)
 \put(0,0){\parbox{0.75\textwidth}{\small\texttt{Contribution to `The Oxford Handbook of Random Matrix Theory', edited by G. Akemann, J. Baik, and P. Di Francesco (Oxford University Press, to be published)}}}
\end{picture}
\vspace*{-4ex}

\begin{center}
{\bf Abstract}
\end{center}
This is a brief overview of RMT applications to quantum or wave chaotic resonance scattering with an emphasis on non-perturbative methods and results.

\sect{Introduction}\label{intro}

Describing bound states corresponding to discrete energy levels in \emph{closed} quantum systems,  one usually addresses properties of Hermitian random matrices (Hamiltonians) $H$. Experimentally, however, one often encounters the phenomenon of chaotic scattering of quantum waves (or their classical analogues) in \emph{open} systems \cite{Kuhl2005a}. The most salient feature of open systems is the set of resonances which are quasibound states embedded into the continuum. The resonances manifest themselves via fluctuating structures in scattering observables. A natural way to address them is via an energy-dependent scattering matrix, $S(E)$, which relates the amplitudes of incoming and outgoing waves. In such an approach, the resonances correspond to the poles of $S(E)$ located (as required by the causality condition) in the lower half-plane of the complex energy plane \cite{Nussenzveig}.

The correspondence becomes explicit in the framework of the Hamiltonian approach \cite{Verbaarschot1985}. It expresses the resonance part of the $S$-matrix via the Wigner reaction matrix $K(E)$ (which is a $M{\times}M$ Hermitian matrix) as follows
\begin{equation}\label{S1}
 S(E) = \frac{1-iK(E)}{1+iK(E)} \,, \qquad K(E) = \textstyle\frac{1}{2}V^{\dag} (E-H)^{-1} V \,,
\end{equation}
where $V$ is an $N{\times}M$ matrix of energy-independent coupling amplitudes between $N$ levels and $M$ scattering channels. By purely algebraic transformations, the above expression can be equivalently represented in terms of the $N{\times}N$ effective non-Hermitian Hamiltonian $\Heff$ of the open system:
\begin{equation}\label{S2}
 S(E) = 1- iV^{\dag} \frac{1}{E-\Heff} V \,, \qquad \Heff = H - \textstyle\frac{i}{2}VV^{\dag}\,.
\end{equation}
The anti-Hermitian part of $\Heff$ has a factorized structure which is a direct consequence of the unitarity condition (flux conservation): $S^{\dag}(E)S(E)=1$ for real $E$. It describes a coupling of the bound states via the common decay channels, thus converting the parential levels into $N$ \emph{complex} resonances $\mathcal{E}_n=E_{n}-\frac{i}{2}\Gamma_{n}$ characterized by energies $E_{n}$ and widths $\Gamma_n>0$. The eigenvectors of $\Heff$  define the corresponding \emph{bi-orthogonal} resonance states (quasimodes). They appear via the residues of the $S$-matrix at the pole positions.

Universal properties of both resonances and resonance states in the chaotic regime are analysed by replacing the actual non-Hermitian Hamiltonian with an RMT ensemble of appropriate symmetry.\footnote{Conventionally,  $H$ is taken from the GOE or GUE labeled by the Dyson's index $\beta=1$ or $\beta=2$, respectively, according to time-reversal symmetry being preserved or broken in the system (GSE, $\beta=4$, is to be taken when spin becomes important).}
In the limit $N\to\infty$ spectral fluctuations on the scale of the mean level spacing $\Delta$ turn out to be universal, i.e. independent of microscopical details such as the particular form of the distribution of $H$ or the energy dependence of $\Delta$. Similarly, the results are also independent of particular statistical assumptions on coupling amplitudes $V_n^c$ as long as $M\ll N$ \cite{Lehmann1995a}. The amplitudes may be therefore equivalently chosen fixed \cite{Verbaarschot1985} or random \cite{Sokolov1989} variables and enter final expressions only via the so-called transmission coefficients
\begin{equation}\label{T}
 T_c\equiv1-|\overline{S}_{cc}|^2=\frac{4\kappa_c}{(1+\kappa_c)^2}\,, \qquad
 \kappa_c=\frac{\pi\|V^c\|^2}{2N\Delta}
\end{equation}
where $\overline{S}$ stands for the average (or optical) $S$ matrix. The set of $T_c$'s is assumed to be the only input parameters in the theory characterising the degree of system openness. The main advantage of such an approach is that it treats on equal footing both the spectral and scattering characteristics of open chaotic systems and is flexible enough to incorporate other imperfections of the system, e.g., disorder due to random scatterers and/or irreversible losses \cite{Fyodorov2005r}.

In fact, the description of wave scattering outlined above can be looked at as an integral part of the general theory of linear dynamic open systems in terms of the input-output approach. These ideas and relations  were developed in system theory and engineering mathematics many years ago, going back to the pioneering works by M. Liv\v{s}ic \cite{Livsic}. A brief description of main constructions and interpretations of the linear open systems approach, in particular, a short derivation of Eq.~(\ref{S2}), can be found in \cite{Fyodorov2000}. In what follows, we overview briefly selected topics on universal statistics of resonances and scattering observables, focusing mainly on theoretical results obtained via non-perturbative methods starting from mid-nineties. For a more detailed discussion, including applications, experimental verifications  as well as references, we refer to the special issue of J. Phys. A \textbf{38}, Number 49 (2005) on `Trends in Quantum Chaotic Scattering' as well as to the recent reviews \cite{Beenakker1997,Guhr1998,Mello1999,Alhassid2000,Mitchell2010}. Finally, it is worth mentioning that
many methods and results discussed below can be extended beyond the universal RMT regime and the use of more ramified random matrix ensembles (e.g. of banded type) allows one to take into account effects of the Anderson localisation. Discussion of those developments goes, however, beyond the remit of the present review, and an interested reader may find more information and further references in \cite{Kottos2002,Fyodorov2003b,Mendez2005,Kottos2005,Ossipov2005,Fyodorov2005r,Weiss2006,Mirlin2006,Monthus2009}.

\sect{Statistics at the fixed energy}\label{distr}

Statistical properties of scattering observables considered at the given fixed value of the scattering energy $E$ can be inferred from the corresponding distribution of $S=S(E)$. The later is known to be uniquely parameterized by the average $S$-matrix and is distributed according to Poisson's kernel \cite{Hua,Mello1985}:
\begin{equation}\label{poisson}
 P_{\overline{S}}(S)  =\frac{1}{V_\beta}\left| \frac{\det[1- \overline{S}^{\dag}\overline{S}]}{
\det[1- \overline{S}^{\dag}S]^2}\right|^{(\beta M +2-\beta)/2},
\end{equation}
where $V_\beta$ is a normalization constant. Although this expression follows from Eq.~(\ref{S2}) in the limit $N\to\infty$, as shown by Brouwer \cite{Brouwer1995}, $P_{\overline{S}}(S)$ can be derived starting from statistical assumptions imposed on $S$ without referencing to the Hamiltonian at all, as it was initially done by Mello et al. \cite{Mello1985}. For a detailed account of the theory see the recent review \cite{Mello1999} and book \cite{Mello2004}.

\subsection{Maximum entropy approach}

 The starting point of that information-theoretical description is an information entropy $\mathcal{S}=-\int d\mu(S)P_{\overline{S}}(S)\ln P_{\overline{S}}(S)$ associated with the probability distribution $P_{\overline{S}}(S)$. The integration here is over the invariant (Haar) measure $d\mu(S)$ which satisfies the symmetry constraints imposed on $S$: unitary $S$ should be chosen symmetric for $\beta=1$ or self-dual  for $\beta=4$.  Due to the causality condition, $S(E)$ is analytic in the upper half of the complex energy plane, implying $\overline{S^k}=\overline{S}^k$ for the positive moments, where the average is assumed to be performed over an energy interval. One can further assume that $S(E)$, for real $E$, is a stationary random-matrix function satisfying the condition of ergodicity, i.e. ensemble averages are equal to spectral averages, the later being determined by the (given) optical scattering matrix $\overline{S}$. These two conditions together yield the so-called \emph{analyticity-ergodicity} requirement
$ \int d\mu(S) S^k P_{\overline{S}}(S) = \overline{S}^k$, $k=1,2,\ldots$. The sought distribution is then assigned to an ``equal a priory'' distribution of $S$, i.e. the one which maximizes the entropy $\mathcal{S}$, subject to the above constraints. The answer turns out to be exactly given by Eq.~(\ref{poisson}) \cite{Mello1985}, with $V_\beta=\int d\mu(S)$ being the volume of the matrix space. Then for any analytic function $f(S)$ its average is given by  the formula  $\int d\mu(S)f(S)P_{\overline{S}}(S) = f(\overline{S})$. Thus the system-specific details are irrelevant, except for the optical $S$-matrix.

The important case of $\overline{S}=0$ (ideal, or perfect coupling) plays a special role, being physically realised when  prompt or direct scattering processes described by $\overline{S}$ are absent. The case corresponds to Dyson's circular ensembles of random unitary matrices $S$ distributed uniformly with respect to the invariant measure (note $P_0(S)=\mathrm{constant}$). Thus statistical averaging amounts to the integration over unitary group,  various methods being developed for this purpose \cite{Mello1985,Brouwer1996}. The general situation of nonzero $\overline{S}$ is much harder for analytical work. In this case, however, one can use the following formal construction
\begin{equation}\label{pois-map}
 S_0 = (t'_1)^{-1} (S-\overline{S})(1-\overline{S}^{\dag}S)^{-1} t_1^{\dag}\,,
\end{equation}
where $S_0$ is drawn from the corresponding circular ensemble and the matrices $t_1$ and $t_1'$ belong to a $2M{\times}2M$ unitary matrix $S_1={r_1\  t_1' \choose t_1 \ r_1'}$ of the same symmetry. Evaluating the Jacobian of the transformation (\ref{pois-map}), one finds \cite{Hua,Friedman1985}  that the distribution of $S$ is exactly given by the Poisson's kernel (\ref{poisson}) provided that we identify $r_1=\overline{S}$. Physically, this can be understood by subdividing the total scattering process into the stage of prompt (direct) response $S_1$ occurring prior to the stage of equilibrated (slow) response $S_0$ \cite{Brouwer1996,Mello1999}, Eq.~(\ref{pois-map}) being a result of the composition of the two processes. In this way one can map the problem  with direct processes to that without and several useful applications were considered in \cite{Gopar1998,Savin2001,Fyodorov2005r,Gopar2008}.

In problems of quantum transport it is often convenient to group scattering channels into $N_1$ `left' and $N_2$ `right' channels  which correspond to the propagating modes in the two leads attached to the conductor (with $M=N_1+N_2$ channels in total). In this way the matrix $S$ acquires the block structure
\begin{equation}\label{S}
  S = \left( \begin{array}{cc} r & t' \\ t & r' \end{array}\right)
    = \left( \begin{array}{cc} u & 0 \\ 0 & v \end{array}\right)
    \left( \begin{array}{cc} -\sqrt{1{-}\tau^T\tau} & \tau^T \\ \tau & \sqrt{1{-}\tau\tau^T} \end{array}\right)
    \left( \begin{array}{cc} u' & 0 \\ 0 & v' \end{array}\right),
\end{equation}
where $r$ ($r'$) is a $N_1{\times}N_1$ ($N_2{\times}N_2$) reflection matrix in the left (right) lead and $t$ ($t'$) is a rectangular matrix of transmission from left to right (right to left). Unitarity ensures that the four Hermitian matrices $tt^{\dag}$, $t't'^{\dag}$, $1-rr^{\dag}$ and $1-r'r'^{\dag}$ have the same set of $n\equiv\mathrm{min}(N_1,N_2)$ nonzero eigenvalues \{$\T_k\in[0,1]$\}. In terms of these so-called transmission eigenvalues the $S$ matrix can be rewritten by employing the singular value decomposition for $t=v\tau u'$ and $t'=u\tau^{T}v'$ yielding the second equality in (\ref{S}). Here $\tau$ is a $N_1{\times}N_2$ matrix with all elements zero except $\tau_{kk}=\sqrt{\T_k}$, $k=1,\ldots,n$, $u$ and $u'$ ($v$ and $v'$) being $N_1{\times}N_1$ ($N_2{\times}N_2$) unitary matrices ($u^*u'=v^*v'=1$ for $\beta=1,4$). For $S$ drawn from one of Dyson's circular ensembles those unitary matrices are uniformly distributed (with respect to the Haar measure), whereas the joint probability density function (JPDF) of the transmission eigenvalues reads \cite{Baranger1994,Jalabert1994}
\begin{equation}\label{jpdf}
  \mathcal{P}_\beta(\{\T_i\}) = \frac{1}{\mathcal{N}_\beta} |\Delta(\{\T_i\})|^{\beta}\prod_{i=1}^n
    \T_i^{\alpha -1}\,, \qquad \alpha\equiv\frac{\beta}{2}(|N_1-N_2|+1)\,.
\end{equation}
Here and below $\Delta(\{x_i\})=\prod_{i<j}(x_i-x_j)$ denotes the Vandermonde determinant, $\mathcal{N}_\beta$ being the normalization constant. Since the Landauer-B\"uttiker scattering formalism allows one to express various transport observables in terms of $T_j$'s \cite{Blanter2000}, Eq.~(\ref{jpdf}) serves as the starting point for performing statistical analysis of quantum transport in chaotic systems, see \cite{Beenakker1997} for a review.

\subsection{Quantum transport and the Selberg integral}

The conductance $g=\mathrm{tr}(tt^{\dag})=\sum_{i=1}^{n}\T_i$ and the (zero-frequency) shot-noise  power $p=\sum_{i=1}^{n}\T_i(1-\T_i)$ are important examples of a linear statistic on transmission eigenvalues. Their mean values are provided by averaging over $t$ or by making use of the mean eigenvalue density (derived recently in \cite{Vivo2008a}). However, the higher moments involve products of $\T_j$'s and require to integrate over the full JPDF. This task can be accomplished by recognizing a profound connection of (\ref{jpdf}) to the celebrated Selberg integral \cite{Savin2006}
which is a multidimensional generalization of Euler's beta-function.\footnote{
 $\int_{0}^{1}\!\mathrm{d}T_1\!\cdots\!\int_{0}^{1}\!\mathrm{d}T_n
 |\Delta(T)|^{2c} \prod_{i=1}^{n} T_i^{a-1}(1-T_i)^{b-1}=
 \prod_{j=0}^{n-1}\frac{\Gamma(1+c+jc)\Gamma(a+jc)\Gamma(b+jc)}{
 \Gamma(1+c)\Gamma(a+b+(n+j-1)c)}$, $\Gamma(x)$ being the gamma function, gives the definition of Selberg integral valid at integer $n\ge1$,  complex $a$ and $b$ with positive real parts, and complex $c$ with $\mathrm{Re\,}c>-\mathrm{min}[\frac{1}{n}$, $\mathrm{Re\,}\frac{a}{n-1}$, $\mathrm{Re\,}\frac{b}{n-1}]$. The constant $\mathcal{N}_{\beta}$ is given by this expression at $a=\alpha$, $b=1$ and $c=\frac{\beta}{2}$. Chapter 17 of Mehta's book gives an introduction into the Selberg integral theory, see \cite{Forrester2008} for its current status.}
The specific structure of its integrand yields a closed set of algebraic relations for the moments $\langle\T_1^{\lambda_1}\cdots\T_m^{\lambda_m}\rangle$ of given order $r=\sum \lambda_i$ with integer parts $\lambda_i\ge0$ \cite{Savin2008}.

As a particular example, we consider the relations for the second order moments:
$
 [\alpha+2+\beta(n-1)] \langle{\T_1^2}\rangle = [\alpha+1+\beta(n-1)]\langle{\T_1}\rangle - \frac{\beta}{2}(n-1)\langle{\T_1\T_2}\rangle
$
and
$
 \langle{\T_1\cdots\T_m}\rangle = \prod_{j=1}^m \frac{\alpha+\beta(n-j)/2}{\alpha+1+\beta(2n-j-1)/2}
$
at $m=2$. These provide us straightforwardly with the well-known exact results for both the average and variance of the conductance (first obtained by different methods in \cite{Baranger1994,Jalabert1994,Brouwer1996}) as well as with the average shot-noise power (hence, Fano factor $\frac{\langle{p}\rangle}{\langle{g}\rangle}$) \cite{Savin2006}:
\begin{equation}
  \langle{g}\rangle = \frac{N_1N_2}{M+1-\frac{2}{\beta}}, \qquad
  \langle{p}\rangle = \langle{g}\rangle \frac{(N_1-1+\frac{2}{\beta})(N_2-1+\frac{2}{\beta}) }{
 (M-2+\frac{2}{\beta})(M-1+\frac{4}{\beta})}\,.
\end{equation}
(We note also an interesting exact relation $\mathrm{var}(g)=\frac{2}{\beta}(N_1N_2)^{-1} \langle{g}\rangle \langle{p}\rangle$.) The approach has been developed further to study full counting statistics of charge transfer \cite{Novaes2007} as well as to obtain explicit expressions  for the skewness and kurtosis of the charge and conductance distributions, and for the shot-noise variance \cite{Savin2008}. With symmetry index $\beta$ entering the Selberg integral as a continuous parameter,  the method allows one to treat all the three ensembles on equal footing and provides a powerful non-perturbative alternative to orthogonal polynomial \cite{Araujo1998,Vivo2008a} or group integration approaches \cite{Brouwer1996,Bulgakov2006}.

It is further possible to combine Selberg integral with the theory of symmetric functions and to develop a systematic approach for computing the moments of the conductance and shot-noise (including their joint moments) of arbitrary order and at any number of open channels \cite{Khoruzhenko2009}. The method is based on expanding powers of $g$ or $p$ (or other linear statistics) in Schur functions $s_\lambda$. These functions are symmetric polynomials
$
  s_{\lambda}(\T) = \det \{\T_i^{\lambda_j+n-j}\}_{i,j=1}^n / \det\{\T_i^{n-j}\}_{i,j=1}^n
$
in $n$ transmission eigenvalues $\T_1, \ldots, \T_n$ indexed by partitions $\lambda$.\footnote{
 A partition is a finite sequence $\lambda=(\lambda_1, \lambda_2, \ldots , \lambda_m)$ of non-negative integers (called parts) in decreasing order  $\lambda_1 \ge \lambda_2 \ge \ldots \ge \lambda_m \ge 0$. The weight of a partition, $|\lambda|$, is the sum of its parts, $|\lambda|=\sum_j \lambda_j$, and the length, $l(\lambda)$, is the number of its non-zero parts.
}
In the group representation theory the Schur functions are the irreducible characters of the unitary group and hence are orthogonal. This orthogonality can be exploited to determine the ``Fourier coefficients'' in the Schur function expansion by integration over the unitary group. The general form of this expansion  reads:
\begin{eqnarray}\label{exp}
  \prod_{i=1}^n \biggl(\sum_{j=-\infty}^{+\infty} a_j{\T_i}^j \biggr)
  = \sum_{\lambda}c_{\lambda}(a) s_{\lambda}(\T),
  \qquad
  c_{\lambda}(a) \equiv \det\bigl\{a_{\lambda_k-k+l}\bigr\}_{k,l=1}^n.\quad
\end{eqnarray}
The summation here is over all partitions $\lambda$ of length $n$ or less, including empty partition for which $s_{\lambda}=1$. The Schur functions can be then averaged over the JPDF (\ref{jpdf})
with the help of integration formulas due to \cite{Hua}. One gets
\begin{eqnarray}\label{mg1aa}
 \langle s_{\lambda} \rangle_{\beta=1} &=& c_{\lambda}\ \prod_{j=1}^{l(\lambda)} \frac{(\lambda_j+N_1-j)!}{(N_1-j)!}\ \frac{(\lambda_j+N_2-j)!}{(N_2-j)!}
 \frac{(M-l(\lambda)-j)!}{(\lambda_j+M-l(\lambda)-j)!}
 \nonumber \\
 &&\times\!\prod_{1\le i\le j\le l(\lambda)} \frac{M+1-i-j}{M+1+\lambda_i+\lambda_j-i-j} \,,
\end{eqnarray}
where the coefficient
$
c_{\lambda} = \frac{\prod_{1\le i<j\le l(\lambda)}(\lambda_i-i-\lambda_j+j)}{\prod_{j=1}^{l(\lambda)}\, (l(\lambda)+\lambda_j-j)!}
$
has been introduced, and
\begin{eqnarray}\label{mg2a}
 \langle s_{\lambda} \rangle_{\beta=2} &=& c_{\lambda} \prod_{j=1}^{l(\lambda)} \frac{(\lambda_j+N_1-j)!}{(N_1-j)!} \frac{(\lambda_j+N_2-j)!}{(N_2-j)!} \frac{(M-j)!}{(\lambda_j+M-j)!}\,,\quad
\end{eqnarray}
in the cases of orthogonal ($\beta=1$)  and unitary ($\beta=2$) symmetry, respectively.

As an application of the method we consider the moments of $\epsilon g+p$. This quantity has a physical meaning of the total noise including both thermal and shot-noise contributions \cite{Blanter2000}. From expansion (\ref{exp}) one finds \cite{Khoruzhenko2009} the $r$-th moment of the total noise given by
\begin{equation}\label{gpexp}
 \langle(\epsilon g + p)^r \rangle = r! \sum_{m=0}^{r} (-1)^m (1+\epsilon)^{r-m}
  \!\!\!\!\sum_{|\lambda|=r+m} f_{\lambda,m} \langle s_{\lambda} \rangle,\ \
\end{equation}
where the second sum is over all partitions of $r{+}m$. Here the expansion coefficient $f_{\lambda,m} = \sum \det\{ \frac {1} {{k_i}!(\lambda_i-i+j-2k_i)!}\}$, where the summation indices $k_i$ run over all integers from 0 to $m$ subject to the constraint $k_1+\ldots +k_{l(\lambda)}=m$. In the limit $\epsilon\to\infty$ Eq.~(\ref{gpexp}) yields all the conductance moments
$\langle{g^r}\rangle = r! \sum_{|\lambda|=r} c_{\lambda} \langle s_{\lambda} \rangle$ (note that $f_{\lambda,0}=c_\lambda$ introduced above) first obtained in \cite{Novaes2008} at $\beta=2$ (see \cite{Osipov2008} for an alternative treatment). In the opposite case of $\epsilon=0$ one gets all the moments of shot-noise, hence all cumulants by well-known recursion relations. Apart from producing lower order cumulants analytically, the method can be used for computing higher order cumulants symbolically by employing computer algebra packages. In particular, one can perform an asymptotic analysis of the exact expressions in the limit $N_{1,2}\gg1$ to make predictions for the leading order term in the $\frac{1}{N}$ expansion of higher-order cumulants \cite{Khoruzhenko2009}.

As concerns the corresponding distribution functions, explicit expressions for the conductance distribution $P_g^{(\beta)}(g)$ are available in particular cases of $n=1,2$ \cite{Mello1999}. At $N_2=K\geq N_1=1$, Eq.~(\ref{jpdf}) readily gives $P_{g,n=1}^{(\beta)}(g)=(\beta K/2)g^{\beta K/2-1}$, $0<g<1$. In the case of $N_1=2$ and arbitrary $N_2=K\ge2$, one can actually find an exact result at any positive integer $\beta$  \cite{Khoruzhenko2009} which reads:
$ P_{n=2}^{(\beta)}(g) = Kg^{\beta K-1} [ X_1 - (-1)^{(\beta K-1)/2} X_2\, \Theta(g-1)
 \sum_{j=0}^{\beta} {\beta \choose j} B_{1-g} (\frac{\beta}{2}(K-1)+j, 1-\beta K)]
$
for $0<g<2$ and zero otherwise. Here $B_{z}(a,b)$ stands for the incomplete beta-function,
$X_1 = \frac{\Gamma[\beta(K+1)/2+1] \Gamma(\beta K/2) }{ \Gamma(\beta/2)\Gamma(\beta K)}$ and
$X_2 = \frac{\Gamma[\beta(K+1)/2+1] }{ \Gamma(\beta)\Gamma[\beta(K-1)/2]}$. As both $N_{1}$ and $N_2$ grow, the distribution of the conductance or, generally, of any linear statistic rapidly approaches a Gaussian distribution \cite{Beenakker1993}. With exact expressions for higher-order cumulants in hand, one can obtain next order corrections to the Gaussian law by making use of the Edgeworth expansion \cite{Blinnikov1998}. Such approximations were shown to be fairly accurate in the bulk even for small channel numbers. Actually, one can derive an explicit representation for the distribution functions in terms of Pfaffians. It works in the whole density support including the spectral edges where the distributions have a power-law dependence. Such asymptotics can be investigated exactly for any $\beta$ and $N_{1,2}$, including powers and corresponding pre-factors, see \cite{Khoruzhenko2009} for details.

\sect{Correlation properties}\label{corr}

The maximum-entropy approach proved to be a success for extracting many quantities important for studying electronic transport in mesoscopic systems, see Chap.~35. At the same time, correlation properties of the $S$-matrix at different values of energy $E$ as well as other spectral characteristics of open systems related to the resonances turn out to be inaccessible in the framework of the maximum-entropy approach because of the single-energy nature of the latter. To address such quantities one has to employ the Hamiltonian approach based on (\ref{S2}). Being supplemented with the powerful supersymmetry technique of ensemble averaging, this way resulted in advances in calculating and studying two-point correlation functions of  $S$-matrix entries and many other characteristics, as discussed below.  However, exact analytical treatment of the higher-order correlation functions (needed, e.g., in the theory of Ericsson fluctuations, see Chap.~2) remains yet an important outstanding problem.

\subsection{$S$-matrix elements}

The energy correlation function of $S$-matrix elements is defined as follows
\begin{equation}\label{Scorr}
C^{abcd}_S(\omega)\equiv\aver{S^{ab*}_{\mathrm{fl}}(E_1)S^{cd}_{\mathrm{fl}}(E_2)}
=\int_0^{\infty}\!\! d t\, e^{2\pi i\omega t} \hat{C}^{abcd}_S(t)\,,
\end{equation}
where $S_{\mathrm{fl}}=S-\overline{S}$ stands for the fluctuating part of the $S$-matrix. Being interested in local fluctuations (on the energy scale of the mean level spacing), it is natural to measure the energy separation in units of $\Delta$. In the RMT limit $N{\to}\infty$, function (\ref{Scorr}) turns out to depend only on the frequency $\omega=\frac{E_2-E_1}{\Delta}$. The Fourier transform $\hat{C}^{abcd}_S(t)$ describes a gradual loss of correlations in time ($\hat{C}^{abcd}_S(t)=0$ at $t<0$ identically, as required by causality). Physically, it is related to the total current through the surface surrounding the scattering centre, thus, a decay law of the open system \cite{Lyuboshitz1978i,Dittes2000}.

The energy dependence becomes explicit if one considers the pole representation of the $S$-matrix which follows from (\ref{S2}): $S^{ab}(E)=\delta^{ab}-i\sum_n \frac{w_n^a\tilde{w}_n^b}{E-\mathcal{E}_n}$. Due to unitarity constraints imposed on $S$, the residues and poles develop nontrivial mutual correlations \cite{Sokolov1989}. For this reason the knowledge of only the JPDF of the resonances $\{\mathcal{E}_n\}$ (considered in the next section) is insufficient to calculate (\ref{Scorr}). The powerful supersymmetry method (see Chap.~7) turns out to be an appropriate tool to perform the statistical average in this case. In their seminal paper \cite{Verbaarschot1985}, Verbaarschot, Weidenm\"uller and Zirnbauer performed the exact calculation of (\ref{Scorr}) at arbitrary transmission coefficients in the case of orthogonal symmetry. The corresponding expression for unitary symmetry was later given in \cite{Fyodorov2005r}. These results are summarized below.

The exact analytic expression for $C^{abcd}_S(\omega)$ can be represented as follows:
\begin{eqnarray}\label{ScorrFT}
 C^{abcd}_S = \delta^{ab}\delta^{cd}T_aT_c\sqrt{(1{-}T_a)(1{-}T_c)} J_{ac} + (\delta^{ac}\delta^{bd}+\delta_{1\beta}\delta^{ad}\delta^{bc})T_aT_bP_{ab}\,.\ \ \
\end{eqnarray}
(The same representation holds for $\hat{C}^{abcd}_S(t)$, so the argument can be omitted.) Here, the $\delta_{1\beta}$ term accounts trivially for the symmetry property $S^{ab} = S^{ba}$ in the presence of time-reversal symmetry. Considering expression (\ref{ScorrFT}) in the energy domain, the functions $J_{ac}(\omega)$ and $P_{ab}(\omega)$  can be generally written as certain expectation values in the field theory (nonlinear zero-dimensional supersymmetric $\sigma$-model). In the $\beta=1$ case of orthogonal symmetry, one has
\begin{equation}\label{SSgoe}
\begin{array}{l}
 J_{ac}(\omega) =  \Bigl\langle
 \Bigl(\sum\limits_{i=1}^{2}\frac{\mu_i}{1+T_a\mu_i}+\frac{2\mu_0}{1-T_a\mu_0}\Bigr)
 \Bigl(\sum\limits_{i=1}^{2}\frac{\mu_i}{1+T_c\mu_i}+\frac{2\mu_0}{1-T_c\mu_0}\Bigr)
 \mathcal{F}_M \Bigr\rangle_{\mu}^{\mathrm{goe}}
\\ \ \\
 P_{ab}(\omega) = \Bigl\langle \Bigl(
 \sum\limits_{i=1}^{2}\frac{\mu_i(1+\mu_i)}{(1+T_a\mu_i)(1+T_b\mu_i)}
 +\frac{2\mu_0(1-\mu_0)}{(1-T_a\mu_0)(1-T_b\mu_0)} \Bigr) \mathcal{F}_M
 \Bigr\rangle_{\mu}^{\mathrm{goe}}
\end{array}
\end{equation}
where $\mathcal{F}^{}_{M}=\prod_c[\frac{(1-T_c\mu_0)^2}{(1+T_c\mu_1)(1+T_c\mu_2)}]^{1/2}$
is the so-called ``channel factor'', which accounts for system openness, and $\langle(\cdots)\rangle_{\mu}^{\mathrm{goe}}$ is to be understood explicitly as
\begin{equation}
\nonumber
 \frac{1}{8} \int_{0}^{\infty}\!d\mu_1 \!\int_{0}^{\infty}\!d\mu_2
 \!\int_{0}^{1}\frac{d\mu_0 \,(1-\mu_0)\mu_0
 |\mu_1-\mu_2|\,e^{i\pi\omega(\mu_1+\mu_2+2\mu_0)} }{
 [(1+\mu_1)\mu_1(1+\mu_2)\mu_2]^{1/2} (\mu_0+\mu_1)^2 (\mu_0+\mu_2)^2 }
 \left(\ldots \right)\,.
\end{equation}
In the $\beta=2$ case of unitary symmetry, the corresponding expressions read:
\begin{equation}\label{SSgue}
\begin{array}{l}
 J_{ac}(\omega) = \Bigl\langle
 \Bigl(\frac{\mu_1}{1+T_a\mu_1}+\frac{\mu_0}{1-T_a\mu_0}\Bigr)
 \Bigl(\frac{\mu_1}{1+T_c\mu_1}+\frac{\mu_0}{1-T_c\mu_0}\Bigr)
 \mathcal{F}_M\Bigr\rangle_{\mu}^{\mathrm{gue}}
\\ \ \\
 P_{ab}(\omega) = \Bigl\langle \Bigl( \frac{\mu_1(1+\mu_1)}{(1+T_a\mu_1)(1+T_b\mu_1)}
 +\frac{\mu_0(1-\mu_0)}{(1-T_a\mu_0)(1-T_b\mu_0)} \Bigr)\mathcal{F}_M\Bigr\rangle_{\mu}^{\mathrm{gue}}
\end{array}
\end{equation}
with $\mathcal{F}_{M}=\prod_c\frac{1-T_c\mu_0}{1+T_c\mu_1}$ and
$\langle(\cdots)\rangle_{\mu}^{\mathrm{gue}} = \int_{0}^{\infty}\!\!d\mu_1\int_{0}^{1}\!d\mu_0
\frac{ \exp\{i2\pi\omega(\mu_1+\mu_0)\} }{ (\mu_1+\mu_0)^{2} }(\ldots)$.

There are several important cases when the above general expressions can be simplified further. We mention first elastic scattering when only a single channel is open (see \cite{Dittes1992} and Eq.~(19) in \cite{Fyodorov2005r}). The profile of the correlation function is strongly non-Lorenzian in this case. This is related to the power-law behavior of the form-factor at large times\footnote{We conventionally measure the time in units of the Heisenberg time $t_H=2\pi/\Delta$ ($\hbar=1)$. The dimensionless time is $t=\mu_0+\frac{1}{2}(\mu_1+\mu_2)$ or $t=\mu_0+\mu_1$ at $\beta=1$ or $\beta=2$, respectively.}: $\hat{C}^{abcd}_S(t)\sim t^{-M\beta/2-2}$  in the general case of $M$ open channels. Such a power-law time decay is a typical characteristic of open chaotic systems  \cite{Lewenkopf1991,Dittes1992}. Formally, it appears due to the dependence $\mathcal{F}_M\sim\prod_c(1+\frac{2}{\beta}T_ct)^{-\beta/2}$ of the channel factor at $t{\gg1}$. Physically, such a behavior can be related to resonance width fluctuations which become weaker as $M$ grows, see Sec.~\ref{sec:decay} below. In the limiting case of the large number $M\gg1$ of weakly open channels, $T_c\ll1$, all the resonances acquire identical escape width $\sum_{c=1}^{M}T_c$, the so-called Weisskopf's width (in units of $\frac{\Delta}{2\pi}$), so that $\mathcal{F}_M$ becomes equal to $e^{-t\sum_{c=1}^{M}T_c}$. As a result, fluctuations in the $S$-matrix are essentially due to those in real energies of the closed counterpart of the scattering system \cite{Verbaarschot1986}, with the final expression in this limit being
\begin{equation}\label{ScorrMT}
 C^{abcd}_S(\omega) =
 \frac{(\delta^{ac}\delta^{bd}+\delta_{1\beta}\delta^{ad}\delta^{bc})T_aT_b
 }{ \sum T_c-2\pi i\omega } +
 \delta^{ab}\delta^{cd}T_aT_c\int_{0}^{\infty}\!\!dt [1-b_2(t)]e^{-(\sum T_c-2\pi i\omega) t}
\end{equation}
where $b_2(t)$ is the canonical two-level RMT form-factor.

As a straightforward application, we mention a prediction for the experimentally accessible elastic enhancement factor $\mathrm{var}(S^{aa})/\mathrm{var}(S^{ab})$, $a\neq b$, which follows from the $S$-matrix correlation function at $\omega=0$. In particular, in the Ericsson's regime of many strongly overlapping resonances ($\sum T_c\gg1$) the first (dominating) term in (\ref{ScorrMT}) yields the well-known Hauser-Feshbach relation. Very recently, the above results have been generalized to the whole crossover regime of gradually broken time-reversal symmetry (GOE--GUE crossover) and also tested in experiments with chaotic microwave billiards \cite{Dietz2009}.

The above formulae  can be also applied to {\it half-scattering} or {\it half-collision} processes (thus {\it decaying} quantum systems). Important examples of such events are represented by photo-dissociation or atomic autoionization  where an absorption of photon excites the quantum system into an energy region (with chaotic dynamics) which allows the subsequent decay. The optical theorem can be used to relate the autocorrelation function of the photo-dissociation cross-sections to  that of $S$-matrix elements (see \cite{Fyodorov1998i} for an alternative approach). Another particular interesting feature is the so-called Fano resonances resulting from an intricate interference of long-time chaotic decays with short-time direct escapes. This can be again described in terms of the $S$-matrix correlations given above, see \cite{Gorin2005} for explicit expressions, detailed discussion and further references.

\subsection{$S$-matrix poles and residues}\label{sec:poles}

\textbf{Resonances.} Within the resonance approximation considered the only singularities of the scattering matrix are its poles $\mathcal{E}_n=E_n-\frac{i}{2}\Gamma_n$. Due to the unitarity constraint their complex conjugates $\mathcal{E}_n^*$ serve as $S$-matrix's zeros. These two conditions yield the following general representation:
\begin{equation}\label{dets1}
 \det{S(E)} =\prod_{k=1}^N \frac{\left(E-{\cal E}_k^*\right)}{\left(E-{\cal E}_k\right)}\,,
\end{equation}
which can be also verified using (\ref{S2}) by virtue of $\det{S(E)}=\frac{\det(E-\Heff^{\dagger})}{\det(E-\Heff)}$. As a result, one can express
the total scattering phaseshift $\phi(E)=\log{\det{S(E)}}$ (or other quantities, e.g, time-delays discussed below) in terms of resonances. Thus statistics of the ${\cal E}_n$'s underlay fluctuations in resonance scattering.

To be able to analyse systematically statistical properties of $N$ eigenvalues ${\cal E}_n$ of $\Heff$, it is natural to start with finding their JPDF $\mathcal{P}_M(\mathcal{E})$. Unfortunately, such a density is known in full generality only at $M=1$. In this case the anti-Hermitian part of $\Heff$ has only one non-zero eigenvalue which we denote $-i\kappa$. Assuming $\kappa>0$ to be non-random, one can find in the GOE case \cite{Fyodorov1999}
\begin{equation} \label{w1a}
 \mathcal{P}_{M=1}^{\mathrm{goe}}(\{\mathcal{E}_i\}) \propto \prod_{k=1}^N
 \frac{e^{-\frac{N}{4}|\mathcal{E}_k|^2} }{ \sqrt{\im\mathcal{E}_k}} \frac{|\Delta(\{\mathcal{E}_i\})|^2}{\prod\limits_{m<n}|\mathcal{E}_m{-}\mathcal{E}^*_n|} \frac{e^{-\frac{N}{4}\kappa^2}}{\kappa^{N/2-1}} \, \delta\biggl(\kappa+\sum_{i=1}^N \im\mathcal{E}_i\biggr)\,,
\end{equation}
whereas in the GUE case the JPDF is given by
\begin{equation} \label{w2a}
 \mathcal{P}_{M=1}^{\mathrm{gue}}(\{\mathcal{E}_i\}) \propto \prod_{k=1}^N e^{-\frac{N}{2}\re\mathcal{E}_k^2}
 |\Delta(\{\mathcal{E}_i\})|^2 \frac{e^{-\frac{N}{2}\kappa^2}}{\kappa^{N-1}}\,\delta\biggl(\kappa+\sum_{i=1}^N \im\mathcal{E}_i\biggr)\,.
\end{equation}
Closely related formulae for randomly distributed $\kappa$ were first derived in \cite{Ullah1969} and, independently, in \cite{Sokolov1989} (see also Ref.~[13] in \cite{Fyodorov2003}). Actually, exploiting a version of the Itzykson-Zuber-Harish-Chandra integral allows one to write the analogue of (\ref{w2a}) at arbitrary $M$ \cite{Fyodorov1999,Fyodorov2003} but the final expression is rather cumbersome.

As typical for RMT problems, the main challenge is to extract the $n$-point correlation functions corresponding to Eq.~(\ref{w1a}) or (\ref{w2a}). The simplest, yet nontrivial statistical characteristics of the resonances is the one-point function, i.e. the mean density. Physically, it can be used to describe the distribution of the resonance widths $\Gamma_n$ in a window around some energy $E$. An important energy scale in such a window is the mean separation $\Delta$ between neighbouring resonances along the real axis.  One can show \cite{Sommers1999} that Eq.~(\ref{w1a}) implies the following probability density for the dimensionless widths $y_n=\pi\Gamma_n/\Delta$:
\begin{equation} \label{w3a}
 \rho_{M=1}^{\mathrm{goe}}(y) = \frac{1}{4\pi}\frac{\partial^2}{\partial y^2} \int_{-1}^1(1-\lambda^2)\,e^{2\lambda y} F(\lambda,y)\,d\lambda
\end{equation}
with
$ F(\lambda,y)=(g{-}\lambda) \int_{g}^\infty\!\!\frac{dp\,\exp(-py) }{ (\lambda-p)^2\sqrt{(p^2-1)(p-g)} }
 \int_{1}^g\!\!\frac{dr\,(p-r)\exp(- r y) }{ (\lambda-r)^2\sqrt{(r^2-1)(g-r)} }
$
where the coupling constant  $g=2/T-1\ge1$ is related to $\kappa$ by Eq.~(\ref{T}). An important feature of this distribution is that it develops an algebraic tail $\propto y^{-2}$ in the case of perfect coupling $g=1$. Such a behaviour was recently confirmed by experimental measurements of resonance widths in microwave cavities \cite{Kuhl2008}.

One can actually derive the distribution of the resonance widths for any finite number $M$ of open channels.  In the GOE case the corresponding expression is rather cumbersome (see Eq.~(3) in \cite{Sommers1999}) but in the GUE case the $M$-channel analogue of (\ref{w3a}) is relatively simple, being given by
\cite{Fyodorov1997}
\begin{equation} \label{w5a}
 \rho_{M}^{\mathrm{gue}}(y) = \frac{(-1)^M}{(M{-}1)!}y^{M-1}\frac{d^M}{dy^M}\left(e^{-yg}\frac{\sinh y}{y}\right)
\end{equation}
for equivalent channels (all $g_c=g$). This formula agrees with the results for chaotic wave scattering in graphs with broken time-reversal invariance \cite{Kottos2000}. Note that in the limit of weak coupling $g\gg 1$  resonances are typically narrow and well-isolated, $\Gamma_n/\Delta\sim T\ll1$. In this regime the above expressions reduce to the standard Porter-Thomas distributions derived long ago by the simple first-order perturbation theory (see discussion and Ref.~[Por65] in Chap.~2).

Actually, for the case of GUE symmetry and the fixed number $M\ll N\to \infty$ of open channels, one can find in full generality not only the distribution of the resonance widths but all the $n$-point correlation functions of resonances ${\cal E}_n$ in the complex plane. Assuming for simplicity a spectral window centered around $\re\mathcal{E}=0$, all correlation functions acquire the familiar determinantal form
\begin{equation} \label{w6a}
\lim_{N\to \infty}\frac{1}{N^{2n}}R_n\left(z_1=N{\cal E}_1,\ldots,z_n=N{\cal E}_n\right)=\det{\left[K(z_i,z^*_j)\right]^n_{j,k=1}}\,,
\end{equation}
where the kernel is given by \cite{Fyodorov1999}
\begin{equation} \label{w6a}
 K(z_1,z^*_2)=F_M^{1/2}(z_1)F_M^{1/2}(z^*_2)\int_{-1}^1 d\lambda\, e^{-i\lambda(z_1-z_2^*)}\prod_{k=1}^M(g_k+\lambda)\,,
\end{equation}
with $F_M(z)=\sum_{c=1}^M \frac{e^{-2|\im(z)|g_c}}{\prod_{s\ne c}(g_c-g_s)}$. In the particular case of equivalent channels the diagonal $K(z,z^*)$ reproduces the mean probability density (\ref{w5a}).

Finally, we briefly describe the behaviour of $S$ matrix poles in the semiclassical limit of many open channels $M\sim N\to \infty$. In such a case the poles form a dense cloud in the complex energy plane characterized by mean density $\rho(z)$ inside the cloud \cite{Haake1992,Lehmann1995a}. The cloud is separated from the real axis by a finite gap.  The gap's width sets another important energy scale, determining the correlation length in scattering fluctuations \cite{Lehmann1995a,Lehmann1995b}.  Considering the $M\to \infty$ limit of $K(z_1,z_2^*)$, one concludes that after appropriate rescaling statistics of resonances inside such a cloud are given by a Ginibre-like  kernel
\begin{equation} \label{w6a}
 \textstyle
 |K(z_1,z^*_2)|=\rho(z)\exp\{-\frac{1}{2}\pi\rho(z)|z_1-z_2|^2\},\quad z=\frac{1}{2}(z_1+z_2)\,.
\end{equation}
This result is expected to be universally valid for non-Hermitian random matrices in the regime of strong non-Hermiticity \cite{Fyodorov2003}, see also Chap.~18.

\noindent\textbf{Nonorthogonal resonance states.} As was mentioned in the introduction, the resonance states (quasimodes) are identified in our approach with  the eigenvectors of the non-Hermitian Hamiltonian
$\Heff$. These eigenvectors form a \emph{bi-orthogonal} rather than orthogonal set, the feature making them rather different from their Hermitian counterparts. More precisely, to any complex resonance energy ${\cal E}_k$ correspond right $|R_k\rangle$  and left $\langle L_k|$ eigenvectors
\begin{equation}\label{eig}
 \Heff|R_k\rangle = \mathcal{E}_k|R_k\rangle \quad \mathrm{and} \quad
 \langle L_k|\Heff =\mathcal{E}_k\langle L_k|
\end{equation}
satisfying the conditions of bi-orthogonality, $\langle{L_k|R_m}\rangle=\delta_{km}$, and completeness, $1=\sum_k|R_k\rangle\langle L_k|$. The non-orthogonality manifests itself via the matrix $\mathcal{O}_{mn}=\langle{L_m|L_n}\rangle\langle{R_n|R_m}\rangle$. In the context of reaction theory such a matrix is known as the Bell-Steinberger non-orthogonality matrix which, e.g., influences branching ratios of nuclear cross-sections \cite{Sokolov1989} and also features in the particle escape from the scattering region \cite{Savin1997}.  In the context of quantum optics $\mathcal{O}_{nn}$ yields the enhancement (Petermann's or excess-noise factor) of the line width of a lasing mode in open resonators whereas $\mathcal{O}_{n\ne m}$  describe cross-correlations between noise emitted in different eigenmodes (see Chap.~36).

This attracted an essential interest in statistical properties of $\mathcal{O}_{nm}$ which were first studied for Ginibre's complex Gaussian ensembles \cite{Chalker1998}. For scattering ensembles,  such a calculation at $\beta=1$  can be successfully performed by re-summing the perturbation theory in the regime of isolated resonances at $M=1$ \cite{Schomerus2000a} or in the opposite case of many open channels,  $M\gg 1$, (strongly overlapping resonances) \cite{Mehlig2001}. The general non-perturbative result for the average density ${\cal O}({\cal E})=\langle\sum_n{\cal O}_{n}\delta({\cal E}_n-{\cal E})\rangle$ of the diagonal parts is available at any $M$ only for $\beta=2$ symmetry \cite{Schomerus2000a}. For the off-diagonal parts, the correlator ${\cal O}({\cal E},{\cal E}')=
\langle\sum_{n\ne m}{\cal O}_{nm}\delta({\cal E}_n-{\cal E})\delta({\cal E}_m-{\cal E'})\rangle$ is known non-perturbatively only for the single-channel case, see \cite{Fyodorov2002ii,Fyodorov2003}.

\subsection{Decay law and width fluctuations}\label{sec:decay}

The decay law of open wave chaotic systems is intimately related to fluctuations of the resonance widths. This important fact can be best quantified by considering the simplest yet non-trivial decay function, namely, the leakage of the norm inside the open system \cite{Savin1997} (we use units of $t_H=\frac{2\pi}{\Delta}=1$ below):
\begin{equation}\label{P(t)}
 P(t) \equiv \frac{1}{N} \Bigl\langle \mathrm{Tr} \Bigl( e^{i\Heff^{\dag}t}e^{-i\Heff t}\Bigr) \Bigr\rangle
  = \frac{1}{N} \Bigl\langle \sum_{n,m} \mathcal{O}_{mn} e^{i(\mathcal{E}^*_n-\mathcal{E}_m)t} \Bigr\rangle\,.
\end{equation}
In the case of the closed system $P(t)=1$ identically at any time. Thus the entire time-dependence of the ``norm-leakage'' $P(t)$ is due to the non-zero widths of the resonance states and their nonorthogonality.

The exact analytic expression for $P(t)$ obtained by supersymmetry calculations turns out \cite{Savin1997,Savin2003i} to be given by that for $P_{ab}(t)$, see Eqs.~(\ref{SSgoe}) and (\ref{SSgue}), where one has to put $T_a=T_b=0$ appearing explicitly in the denominators. The typical behavior $P(t)\sim\prod_c(1+\frac{2}{\beta}T_c t)^{-\beta/2}$. In the so-called ``diagonal approximation'', which neglects nonorthogonality of the resonance states and becomes asymptotically exact at large $t$, $P(t)$ gets simply related to the distribution of resonance widths by the Laplace transform: $P_{\mathrm{diag}}(t)=\frac{1}{N}\langle\sum_n e^{-\Gamma_n t}\rangle=\int_{0}^{\infty}d\Gamma\,e^{-\Gamma t}\rho(\Gamma)$. With Eq.~(\ref{w5a}) in hand, one can then analyse in detail the time evolution in decaying chaotic systems and the characteristic time scales \cite{Savin1997}, see also \cite{Casati1997} for the relevant study. It is worth mentioning further applications of such a consideration to relaxation processes in open disordered conductors \cite{Mirlin2000} and nonlinear random media \cite{Kottos2004}, to chaotic quantum decay in driven biased optical lattices \cite{Glueck2002} and quasi-periodic structures \cite{Kottos2005}, and also in the context of electromagnetic pulse propagation in disordered media \cite{Chabanov2003,Skipetrov2006}.

\sect{Other characteristics and applications}

\subsection{Time-delays}\label{sec:td}

Time delay of an almost monochromatic wave packet is conventionally described via the Wigner-Smith matrix $Q(E) = -i\hbar S^{\dag} dS/dE$, see e.g. \cite{Fyodorov1997,deCarvalho2002} for introduction and historical references. This matrix appears in many applications, e.g., describing charge response and charge fluctuations in chaotic cavities \cite{Buettiker2005}. In the considered resonance approximation of the energy-independent $V$ the matrix $Q(E)$ can be equivalently represented as follows ($\hbar=1$) \cite{Sokolov1997}:
\begin{equation}\label{Q}
  Q(E) = V^{\dag} \frac{1}{(E-\Heff)^{\dag}} \frac{1}{E-\Heff} V.
\end{equation}
This gives the meaning to the matrix element $Q_{ab}$ as the overlap of the internal parts $(E-\Heff)^{-1}V$ of the scattering wave function in the incident channels $a$ and $b$. In particular, the diagonal entries $Q_{cc}$ can be further interpreted as mean time delays in the channel $c$. Their sum $\tau_W=\frac{1}{M}\mbox{Tr} Q$ can be related to the energy-derivative of the total scattering phase, $\tau_W=-\frac{i}{M}\frac{d}{d E}\log{\det{S}}$, and is known as the Wigner time delay. In view of (\ref{dets1}), this quantity is solely determined by the complex resonances: $\tau_W(E)=\frac{1}{M}\sum_{n}\frac{\Gamma_n}{(E-E_n)^2+\Gamma^2_n/4}$.
Its average value is therefore given by the mean level density, $\langle\tau_W\rangle=2\pi/M\Delta$. Fluctuations of $\tau_W$ around this value can be quantified in terms of the autocorrelation function of the fluctuating parts, $\tau_W^{\mathrm{fl}}=\tau_W-\langle\tau_W\rangle$. The exact expression for $\beta=1$ \cite{Lehmann1995b} and $\beta=2$ \cite{Fyodorov1997} cases are given by
\begin{equation}\label{td1}
 \frac{ \langle\tau_W^{\mathrm{fl}}(E-\frac{\Delta}{2}\omega) \tau_W^{\mathrm{fl}}(E+\frac{\Delta}{2}\omega)\rangle }{ \langle\tau_W\rangle^2} = \left\{ \begin{array}{l}
  2\re\left\langle (2\mu_0+\mu_1+\mu_2)^2\right\rangle_\mu^{\mathrm{goe}} \\
  \re\left\langle (\mu_0+\mu_1)^2 \right\rangle_\mu^{\mathrm{gue}}
 \end{array} . \right.
\end{equation}
(The same shorthand as in Eqs.~(\ref{SSgoe})--(\ref{SSgue}) has been used.) Actually, the corresponding result for the whole GOE-GUE crossover is also known \cite{Fyodorov1997i}.

As it is clear from definition (\ref{Q}), the time-delay matrix at a given energy $E$ is in essence a two-point object. Therefore, its distribution cannot be obtained by methods of Sec.~\ref{distr} and requires the knowledge of the distribution functional of $S(E)$. The solution to this problem for the case of ideal coupling, $\overline{S}=0$, was given by Brouwer et al. \cite{Brouwer1997}, who found the time-delay matrix $Q$ to be statistically independent of $S$. Considering the eigenvalues $q_1,\ldots,q_M$ of $Q$ (known as the {\it proper} time delays), the JPDF of their inverses $\tilde{\gamma}_a=(\frac{2\pi}{\Delta})q_a^{-1}$ was found to be given by the Laguerre ensemble as follows \cite{Brouwer1997}:
\begin{equation}\label{td2}
 {\cal P}(\tilde{\gamma}_1,\ldots,\tilde{\gamma}_M)\propto \prod_{a<b}|\tilde{\gamma}_a-\tilde{\gamma}_b|^{\beta}\prod_{a=1}^M\tilde{\gamma}_a^{\beta M/2}
 e^{-\beta\tilde{\gamma}_a/2}.
\end{equation}
The eigenvalue density can be computed by methods of orthogonal polynomials, explicit results being derived for $M\gg1$ at any $\beta$ and for $\beta=2$ at any $M$. In the latter case, one can use (\ref{td2}) for further  extracting the  distribution of $Q_{aa}$ and, finally, that of the Wigner time delay $\tau_W$ \cite{Savin2001} (this paper also provides those distributions at $\beta=1,4$ and $M=1,2$). Representation (\ref{Q}) together with the supersymmetry technique enables us to generalise the above results for the distribution of proper time-delays to the general case of non-ideal coupling \cite{Sommers2001}. In this case, the distribution of the so-called partial time-delays (related to $Q_{cc}$) is actually known in the whole GOE-GUE crossover \cite{Fyodorov1997i}. Finally, we note that in the special case $M=1$ there exists a general relation between the statistics of the (unique in this case) time-delay and that of eigenfunction intensities in the closed counterpart of the system \cite{Ossipov2005}. The RMT predicts that such intensities are distributed for $\beta=1,2,4$ according to the Porter-Thomas distributions, more complicated formulas being available for the $\beta=1$ to $\beta=2$ crossover \cite{Sommers1994,Falko1994}.  In this way one can easily recover, e.g., the above distribution of the time-delay in the crossover regime. The most powerful application of such a relation is however beyond the standard RMT for systems
exhibiting Anderson localisation transition, see \cite{Ossipov2005,Mendez2005,Mirlin2006}.

\subsection{Quantum maps and sub-unitary random matrices}

The scattering approach can be easily adopted to treat open dynamical systems with discrete time, i.e. open counterparts of the so-called area-preserving chaotic maps.  The later are usually represented by unitary operators which act on Hilbert spaces of finite large dimension $N$, being often referred to as evolution, scattering or Floquet operators, depending on the given physical context. Their eigenvalues (eigenphases) consist of $N$ points on the unit circle and conform statistically quite accurately the results obtained for Dyson's circular ensembles, see e.g. \cite{Glueck2002,Ossipov2003,Jacquod2003,Schomerus2009} for diverse models and physical applications. A general scattering approach framework for such systems was developed in \cite{Fyodorov2000} and we mention its gross features below.

For a closed linear system characterized by a wavefunction $\Psi$  the ``stroboscopic'' dynamics amounts to a linear unitary map such that $\Psi(n+1)=\hat{u}\Psi(n)$. The unitary evolution operator $\hat{u}$ describes the inner state domain decoupled both from input and output spaces. Then a coupling that makes the system open must convert the evolution operator $\hat{u}$ to a contractive operator $\hat{A}$ such that $1-\hat{A}^{\dagger}\hat{A}\ge 0$. The equation $\Psi(n+1)=\hat{A}\Psi(n)$ describes now an irreversible decay of any initial state $\Psi(0)\ne 0$ when an input signal is absent. On the other hand, assuming a nonzero input and zero initial state $\Psi(0)=0$, one can relate the (discrete) Fourier-transforms of the input and output signals at a frequency $\omega$ to each other by a $ M\times M$  unitary scattering  matrix $\hat{S}(\omega)$ as follows:
\begin{equation} \label{S-discr}
 \hat{S}(\omega)=\sqrt{1-\hat{\tau}^{\dagger}\hat{\tau}}-\hat{\tau}^{\dagger}
 \frac{1}{e^{-i\omega}-\hat{A}}\hat{u}\hat{\tau}\,, \qquad \hat{A}=\hat{u} \sqrt{1-\hat{\tau}\hat{\tau}^{\dagger}}\,,
\end{equation}
where $\hat{\tau}$ is a rectangular $N\times M$ matrix with $M\le N$ nonzero entries $\tau_{ij}=\delta_{ij}\tau_j$, $0\le \tau_i\le 1$. This formula is a complete discrete-time analogue of Eq.~(\ref{S2}). In particular, one can straightforwardly verify unitarity and show that
\begin{equation}\label{dets2}
 \det{\hat{S}(\omega)}=e^{-i\omega N}
 \frac{\det{\left(\hat{A}^{\dagger}-e^{i\omega}\right)}}
 {\det{\left(e^{-i\omega}-\hat{A}\right)}}
 = e^{-i\omega N}\prod_{k=1}^N
 \frac{\left(z_k^{*}-e^{i\omega}\right)}{\left(e^{-i\omega}-z_k\right)},
\end{equation}
where $z_k$  stand for the complex eigenvalues of the matrix $\hat{A}$ (note $|z_k|<1$). This relation is an obvious analogue of Eq.(\ref{dets1}) and gives another indication of $z_k$ playing the role of resonances for the discrete time systems.

Generic features of quantized maps with chaotic inner dynamics are emulated by choosing $\hat{u}$ from one of the Dyson circular ensembles. By averaging Eq.~(\ref{S-discr}) over $\hat{u}$, one easily finds $\hat{\tau}^{\dagger}\hat{\tau}=1-|\langle\hat{S}\rangle |^2 $. Therefore, $M$ eigenvalues $T_a=\tau_a^2\le1$ of $\hat{\tau}^{\dagger}\hat{\tau}$ play the familiar role of transmission coefficients. In the particular case of all $T_{a}=1$ (ideal coupling), the non-vanishing eigenvalues of the matrix
$\hat{A}$ coincide with those of a $(N{-}M){\times}(N{-}M)$ subblock of $\hat{u}$. Complex eigenvalues of such ``truncations'' of random unitary matrices were first studied analytically in \cite{Zyczkowski2000}. The general ensemble of $N{\times}N$ random contractions $\hat{A}=\hat{u}\sqrt{1-\hat{\tau}\hat{\tau}^{\dagger}}$ was studied in \cite{Fyodorov2003} starting from the following probability measure in the matrix space ($d\hat{A}=\prod d{\re A}_{ij}d{\im A}_{ij}$):
\begin{equation}\label{0}
 \mathcal{P}(\hat{A})d\hat{A} \propto \delta(\hat{A}^{\dagger}\hat{A}-\hat{G}) d\hat{A}\,,\quad
 \hat{G}\equiv {\bf 1} -\hat{\tau}\hat{\tau}^{\dagger}\,,
\end{equation}
Equation (\ref{0}) describes an ensemble of subunitary matrices ${\hat A}$ with given singular values. The question of characterizing the locus of complex eigenvalues for a general matrix with prescribed singular values is classical, and the recent paper \cite{Wei2008} provided a kind of statistical answer to that question.

\subsection{Microwave cavities at finite absorption}

The above consideration is restricted to the idealization neglecting
absorption. The latter is almost invariably present, being usually
seen as a dissipation of incident power evolving exponentially in
time. In the approximation of \emph{uniform} absorption all the
resonances acquire in addition to their escape widths $\Gamma_n$ one
and the same absorption width
$\Gamma_{\mathrm{abs}}\equiv\gamma\frac{\Delta}{2\pi}>0$.
Operationally it is equivalent to a purely imaginary shift of the
scattering energy $E\to E+\frac{i}{2}\Gamma_{\mathrm{abs}}\equiv
E_\gamma$. As a result, the $S$-matrix correlation function in
presence of the absorption is related to that without by replacing
$\omega\to\omega+i\gamma/2\pi$ in (\ref{Scorr}). In the time domain,
the corresponding form-factor then simply acquires the additional
decay factor $e^{-\gamma t}$ (in this way the absorption strength
$\gamma$ can be extracted experimentally, e.g., for the microwave
cavities \cite{Schaefer2003}).

Distribution functions undergo, however, more drastic changes at finite absorption, as the scattering matrix $S(E_\gamma)$ becomes subunitary.
The mismatch between incoming and outgoing fluxes can be quantified by the reflection matrix
\begin{equation}\label{R}
 R = S(E_\gamma)^{\dagger}S(E_\gamma) = 1- \Gamma_{\mathrm{abs}} Q(E_\gamma)\,.
\end{equation}
This representation, with $Q(E_\gamma)$ from (\ref{Q}), is valid at arbitrary $\Gamma_{\mathrm{abs}}$ \cite{Savin2003i}.  The so-called reflection eigenvalues of $R$ play an important role for the description of thermal emission from random media, as discovered by Beenakker, who also computed their distribution at perfect coupling ($T=1$) (see Chap.~36.3). The above connection to the time-delay matrix at finite absorption allows one to generalize this result to the case of arbitrary coupling, $T\le1$ \cite{Savin2003i}.

In terms of quantum mechanics the quantity $1-R_{aa}$ has the meaning of probability of no return to the incident channel. For a single channel scattering without absorption it should be identically zero, but finite absorption induces a ``unitary deficit'' $1-R_{aa}>0$ which can be used as a sensitive probe of the chaotic scattering. The quantity starts to play even more important role in a single-channel scattering from systems with intrinsic disorder. Indeed, for such systems the phenomenon of {\it spontaneous} breakdown of $S$-matrix unitarity was suggested as one of the most general manifestations of the Anderson localisation transition of waves in random media, see \cite{Fyodorov2003b} for further discussion,
and \cite{Kottos2002} for somewhat related ideas.

In the context of statistical electromagnetics, the matrix $Z=iK(E_\gamma)$ has the meaning of the normalized cavity impedance \cite{Hemmady2005}. In view of (\ref{S1}), the joint distribution function of its real and imaginary parts fully determines the distribution of (non-unitary) $S$. It turns out that there exists a general relation (fluctuation-dissipation like) between this function and the energy autocorrelation function of resolvents of $K$ at zero absorption \cite{Savin2005}. This relation enables us further to compute the exact distributions of complex impedances and of reflection coefficients $|S_{cc}|^2$ at any $M$ and for the whole GOE-GUE crossover \cite{Fyodorov2005r}, and to explain the relevant experimental data \cite{Kuhl2005,Zheng2006,Lawniczak2008}.
The corresponding study for the case of transmission is still an open problem.

Quite often, however, an approximation of uniform absorption may break down, and one should take into account \emph{inhomogeneous} (or localized-in-space) losses which result in different broadening of different modes. The latter are easily incorporated in the model by treating them as if induced by additional scattering channels, see e.g. \cite{Rozhkov2003,Savin2006e,Poli2009} for relevant applications.



\end{document}